# Proton Spin Structure from Lattice QCD

M. Fukugita$^{a)\dagger}$, Y. Kuramashi$^{b)}$, M. Okawa$^{b)}$, A. Ukawa$^{c)}$

*Yukawa Institute, Kyoto University$^{a)}$*
*Kyoto 606, Japan*

*National Laboratory for High Energy Physics(KEK)$^{b)}$*
*Tsukuba, Ibaraki 305, Japan*

*Institute of Physics, University of Tsukuba$^{c)}$*
*Tsukuba, Ibaraki 305, Japan*

**Abstract**

A lattice QCD calculation of the proton matrix element of the flavor singlet axial-vector current is reported. Both the connected and disconnected contributions are calculated, for the latter employing the variant method of wall source without gauge fixing. From simulations in quenched QCD with the Wilson quark action on a $16^3 \times 20$ lattice at $\beta = 5.7$ (the lattice spacing $a \approx 0.14$fm), we find
$\Delta\Sigma = \Delta u + \Delta d + \Delta s = +0.638(54) - 0.347(46) - 0.109(30) = +0.18(10)$
with the disconnected contribution to $\Delta u$ and $\Delta d$ equal to $-0.119(44)$, which is reasonably consistent with the experiment.

$^\dagger$ Also at *Institute for Advanced Study, Princeton, NJ 08540, U. S. A.*

The flavor singlet axial-vector matrix element of proton has been widely discussed in recent years. The interest initially arose from the EMC data[1] for the spin-dependent proton structure function $g_1$ which, taken together with earlier SLAC data[2], apparently indicated that the fraction of proton spin carried by quarks has a small value $\Delta\Sigma = \Delta u + \Delta d + \Delta s = 0.12(17)$ and that the strange quark contribution is unexpectedly large and negative $\Delta s = -0.19(6)$[1]. New experiments have since been performed with proton[3, 4], deuteron[4, 5] and neutron[6] targets. Combined reanalyses of these data using $g_A = \Delta u - \Delta d = 1.2573(28)$ and $g_8 = \Delta u + \Delta d - 2\Delta s = 0.601(38)$[7], with the aid of three-loop perturbative QCD calculations for the structure functions[8], have recently been carried out[9, 10], reporting for the matrix elements renormalized at $\mu^2$,

$$\begin{aligned}\Delta\Sigma &= \Delta u + \Delta d + \Delta s \\ &= \begin{cases} 0.83(3) - 0.43(3) - 0.10(3) = 0.31(7), & \mu^2 = 10\text{GeV}^2[9] \\ 0.832(15) - 0.425(15) - 0.097(18) = 0.31(4), & \mu^2 = \infty[10]. \end{cases}\end{aligned} \quad (1)$$

where the first analysis[9] employed the Bjorken sum rule to fix the value of the strong coupling constant $\alpha_s$ while the world average for $\alpha_s$ was used in the second analysis[10].

In this article we report[11] on a quenched lattice QCD calculation of the proton matrix elements of axial-vector current for $u$, $d$ and $s$ quarks including both connected and disconnected contributions. A serious technical obstacle in such a calculation has been a reliable estimate of the disconnected part, which we are now able to overcome with the variant of the method of wall sources[12]. An exploratory study employing a random $Z(2)$ source was previously made in Ref. [13]. Majority of lattice QCD calculations to date, however, attempted to evaluate $\Delta\Sigma$ from the proton matrix element of topological charge density[14, 15, 16]. The use of quenched approximation is not valid in this approach due to the degeneracy of $\eta'$ and $\pi$[14]. It also turned out that data generated in quenched QCD[14, 16] are too noisy to extract the matrix element. From a full QCD calculation with four flavors of Kogut-Susskind quarks, the authors of Ref. [15] reported $\Delta\Sigma = 0.18(2)$. This is a difficult calculation and the quality of their raw data does not seem to be as good as the error they quoted indicates. A direct calculation of axial-vector current matrix elements is superior in that the problem due to $\eta'$ is absent and that the contribution of connected and disconnected contributions for each quark flavor and their quark mass dependence can be examined.

Let us define $<p_s|\bar{q}\gamma_3\gamma_5 q|p_s> = s \cdot \Delta q$ with $|p_s>$ the proton state at rest with the spin projection in the $z$ direction equal to $s/2$. To extract $\Delta q$ we calculate the ratio of the three-point function of the proton and axial-vector current to the proton two-point function, each projected onto the zero momentum state,

$$R(t) = \frac{\sum_{s=\pm} s \cdot <p_s(t) \sum_{t' \neq 0} \bar{q}\gamma_3\gamma_5 q(t') \bar{p}_s(0)>}{\sum_{s=\pm} <p_s(t)\bar{p}_s(0)>} \xrightarrow{\text{large } t} \text{const.} + Z_A^{-1} \Delta q\, t, \qquad (2)$$

with $Z_A$ the lattice renormalization factor for the axial-vector current. The connected amplitude can be calculated by the conventional source method[17]. To handle the disconnected piece we employ quark propagators $G(\mathbf{n}, t)$ evaluated with unit source at every space-time site ( except for the $t = 0$ time slice to avoid mixing with the proton valence quark propagator) without gauge fixing[12]. The product of the nucleon propagator and $\sum_{(\mathbf{n}, t \neq 0)} \text{Tr}[\gamma_3\gamma_5 G(\mathbf{n}, t)]$ equals the disconnected amplitude up to gauge-variant non-local terms which cancel out in the average over gauge configurations.

Our calculation is carried out for the Wilson quark action in quenched QCD at $\beta = 5.7$ on a $16^3 \times 20$ lattice. The $u$ and $d$ quarks are assumed to be degenerate with $K_u = K_d \equiv K_q$, while the strange quark is assigned a different hopping parameter $K_s$. We analyzed 260 configurations for the hopping parameters $K_q, K_s = 0.160, 0.164$ and $0.1665$, generated with the single plaquette action separated by 1000 pseudo-heat bath sweeps. In order to avoid contaminations from the negative-parity partner of the proton propagating backward in time, we employ the Dirichlet boundary condition in the temporal direction for quark propagators. We use the relativistic proton operator, fixing gauge configurations on the $t = 0$ time slice to the Coulomb gauge in order to enhance proton signals. Let us emphasize that the gauge fixing is limited to the $t = 0$ time slice; the rest of the lattice is left unfixed. Thus the issue of contribution of gluons to proton spin through gauge non-invariant Chern-Simons current[18] does not arise in our case.

In Table 1 we list the results for hadron masses obtained by a single exponential fit of propagators over $6 \leq t \leq 12$. Errors in this table and below are estimated by the single elimination jackknife procedure.

We plot the connected contribution of $u$ and $d$ quark to the ratio $R(t)$ in Fig. 1 (a) and (b) for the case of $K_q = 0.164$. Very clean signals with the linear behavior up to $t \approx 14$ are observed, with the sign of the slope consistent with expectations from quark models which predict $\Delta u = 4/3$ and $\Delta d = -1/3$ in the static limit. For the disconnected contribution the quality of our data is not quite good, in spite of reasonably high statistics of the simulation ( Fig. 1 (c) ). A region showing a linear dependence is limited to $t \approx 5 - 10$ and errors grow rapidly with increasing $t$; For $t \geq 12$ signals are lost into a large noise. Nonetheless, we can still observe the negative value of the sea quark contribution to the axial-vector matrix element including that of strange quark, which is not predictable in quark models. To extract the axial-vector matrix elements, we fit the data for $R(t)$ to the linear form (2) with the fitting range chosen to be $5 \leq t \leq 10$. Changing the fitting range to $5 \leq t \leq 11$ or $6 \leq t \leq 10$ increases the value of fitted slope by $30 - 40\%$ with a roughly proportional increase of error. (Including larger values of $t$ does not seem reasonable due to a departure from a linear behavior for $t \geq 11$ and a rapid loss of signals.)

Results are corrected by the tadpole-improved renormalization factor given

by[19]

$$Z_A = (1 - 0.31\alpha_s)\left(1 - \frac{3K}{4K_c}\right), \qquad (3)$$

where we use $\alpha_{\overline{MS}}(1/a) = 0.2207$ for $\alpha_s$. We should note that the flavor singlet axial-vector current requires an additional lattice-to-continuum divergent renormalization from diagrams containing the triangle anomaly diagram. We leave out this factor since the explicit form of this contribution which starts at two-loop order has not been computed yet. The results for $\Delta u$, $\Delta d$ and $\Delta s$ are tabulated in Table 2.

We present the axial-vector matrix element for $u$, $d$ and $s$ quarks in Fig. 2 as functions of the bare $u$ and $d$ quark mass $m_q a = (1/K_q - 1/K_c)/2$ using $K_c = 0.1694$. For the strange quark the values represent result of an interpolation to the physical strange quark mass corresponding to $m_K/m_\rho = 0.64$ as explained below. As we already remarked the disconnected contributions (filled symbols) are negative. Furthermore their magnitude is small, albeit increasing slightly toward the chiral limit. The connected contributions exhibit an opposite trend of a slight decrease.

We calculate the physical values of matrix elements in the following way. For $u$ and $d$ quarks, we estimate the sum of disconnected and connected contributions by first combining the two contributions in the ratio $R(t)$ and then fitting the result to the linear form (2) over $5 \leq t \leq 10$ for each $K_q$. The fitted values are extrapolated linearly to the chiral limit $m_q = 0$. For the strange quark contribution similar extrapolations to $m_q = 0$ are made for each $K_s$, whose results in turn are fitted to a linear function of the strange quark mass $m_s a = (1/K_s - 1/K_c)/2$. The physical strange quark mass is estimated by generalizing the relation $m_\pi^2 a^2 = 2.710(25) m_q a$, obtained from the hadron mass results in Table 1, to $m_K^2 a^2 = 2.710(25)(m_q a + m_s a)/2$ and using the experimental ratio $m_K/m_\rho = 0.64$, which gives $m_s a = 0.0826(22)$ or $K_s = 0.1648$. This analysis, as summarized in Table 3, yields for the quark contribution to proton spin,

$$\Delta\Sigma = \Delta u + \Delta d + \Delta s = +0.638(54) - 0.347(46) - 0.109(30) = +0.18(10), \quad (4)$$

These values, notably the sign and magnitude of the strange quark contribution, show a reasonable agreement with the phenomenological estimate (1)[10].

Table 3 also lists the flavor non-singlet matrix elements $F_A = (\Delta u - \Delta s)/2$ and $D_A = (\Delta u - 2\Delta d + \Delta s)/2$ where the strange quark mass is interpolated to the physical value. In the chiral limit for $u$ and $d$ quarks $m_q = 0$ we find $F_A = 0.382(18)$ and $D_A = 0.607(14)$, which implies $g_A = F_A + D_A = 0.985(25)$ and $F_A/D_A = 0.629(33)$. Compared to the experimental values $g_A = 1.2573(28)$ and $F_A/D_A = 0.586(19)$[7], the ratio shows a good agreement while the magnitude of $F_A$ and $D_A$ are about 25% smaller. Previous quenched results at $\beta = 6.0$[20, 21] and with the $\sqrt{3}$-blocked Wilson action[22], and for full QCD[20] at $\beta = 5.4 - 5.6$, with the lattice spacing of $a \approx 0.15 - 0.1$fm, yield similar results if analyzed with the same renormalization factor as we employed.

Possible sources of systematic errors in our results are scaling violation effects due to a fairly large lattice spacing $a \approx 0.14$fm of our simulation at $\beta = 5.7$ and uncertainties in the perturbative estimate of the renormalization factor (3). The small values of flavor non-singlet couplings compared to experiment by about 25%, possibly arising from these uncertainties, suggest that our result for $\Delta\Sigma$ might be underestimating the continuum value by a similar magnitude. The use of quenched QCD might also cause systematic errors. We note that the lack of two-loop calculation for the flavor singlet lattice-to-continuum renormalization factor makes it difficult to specify the scale at which $\Delta\Sigma$ is evaluated, although we expect the scale dependence to be weak being a two-loop effect. While these points should be examined in future studies, we feel that the encouraging result we found points toward an eventual resolution of the spin crisis issue within lattice QCD.

Let us finally note that a result similar to ours has recently been reported at the Lattice '94 Symposium[23]. Employing the $Z(2)$ noise method for evaluating the disconnected contribution on 24 configurations for an $16^3 \times 24$ lattice at $\beta = 6.0$ in quenched QCD, the authors found
$\Delta\Sigma = \Delta u + \Delta d + \Delta s = +0.78(7) - 0.42(7) - 0.13(6) = +0.22(9)$.

## Acknowledgements

Numerical calculations for the present work have been carried out on HITAC S820/80 at KEK. This work is supported in part by the Grants-in-Aid of the Ministry of Education (Nos. 06NP0601, 05640363, 06640372, 05-7511).

# References

[1] J. Ashman *et al.*, Phys. Lett. **B206**, 364 (1988); Nucl. Phys. **B328**, 1 (1989).

[2] V. W. Hughes *et al.*, Phys. Lett. **B212**, 511 (1988); G. Baum *et al.*, Phys. Rev. Lett. **51**, 1135 (1983).

[3] Spin Muon Collaboration, D. Adams *et al.*, Phys. Lett. **B329**, 399 (1994).

[4] E143 Collaboration, J. McCarthy, in *Proc. of the 27th Int. Conf. on High Energy Physics* (Glasgow, July 1994).

[5] Spin Muon Collaboration, B. Adeva *et al.*, Phys. Lett. **B302**, 533 (1993).

[6] E142 Collaboration, P. L. Anthony *et al.*, Phys. Rev. Lett. **71**, 959 (1993).

[7] Particle Data Group, Phys. Rev. D**50**, 1173 (1994); S. Y. Hsueh *et al.*, Phys. Rev. D**38**, 2056 (1988).

[8] S. A. Larin *et al.*, Phys. Rev. Lett. **66**, 862 (1991); Phys. Lett. **B259**, 345 (1991); preprint CERN-TH.7208/94 (1994).

[9] J. Ellis and M. Karliner, Phys. Lett. **B341**, 397 (1995).

[10] G. Altarelli and G. Ridolfi, preprint CERN-TH.7415/94 (1994).

[11] A summary of our results has been presented at the Lattice '94 Symposium, M. Fukugita, Y. Kuramashi, M. Okawa and A. Ukawa, hep-lat/9412025 (1994) (to appear in *Proc. of the 1994 Int. Symp. on Lattice Field Theory* (Bielefeld, September 1994)).

[12] Y. Kuramashi, M. Fukugita, H. Mino, M. Okawa and A. Ukawa, Phys. Rev. Lett. **71**, 2387 (1993); Phys. Rev. Lett. **72**, 3448 (1994).

[13] S.-J. Dong and K. -F. Liu, Nucl. Phys. **B**(Proc. Suppl.) **30**, 487 (1993).

[14] R. Gupta and J. E. Mandula, Phys. Rev. D**50**, 6931 (1994) and references therein.

[15] R. Altmeyer *et al.*, Nucl. Phys. **B**(Proc. Suppl.)**30**, 483 (1993); Phys. Rev. D**49**, R3087 (1994)

[16] B. Allés *et al.*, Phys. Lett. **B336**, 248 (1994).

[17] C. Bernard *et al.*, in *Lattice Gauge Theory: A Challenge in Large-Scale Computing*, eds. B. Bunk *et al.* (Plenum, New York, 1986); G. W. Kilcup *et al.*, Phys. Lett. **164B**, 347 (1985).

[18] See, *e.g.,* R. D. Carlitz *et al.*, Phys. Lett. **B214**, 229 (1988); A. V. Manohar, Phys. Rev. Lett. **66**, 289 (1991).

[19] G. P. Lepage and P. B. Mackenzie, Phys. Rev. D**48**, 2250 (1993).

[20] R. Gupta *et al.*, Phys. Rev. D**44**, 3272 (1991).

[21] K. F. Liu *et al.*, Phys. Rev. D**49**, 4755 (1994).

[22] S. Güsken *et al.*, Phys. Lett. **B227**, 266 (1989).

[23] S.-J. Dong and K. -F. Liu, preprint UK/94-07 (hep-lat/9412059) (1994) (to appear in *Proc. of the 1994 Int. Symp. on Lattice Field Theory* (Bielefeld, September 1994)).

# Figure Captions

Fig. 1 Connected and disconnected contribution to $R(t)$ for $u$ and $d$ quarks at $\beta = 5.7$ and $K_q = 0.164$ on an $16^3 \times 20$ lattice.

Fig. 2 Axial-vector matrix elements for $u$, $d$ and $s$ quark as a function of degenerate $u$ and $d$ quark mass. Strange quark mass is interpolated to the physical value.

# Tables

Table 1: Hadron mass results on an $16^3 \times 20$ lattice at $\beta = 5.7$ obtained with 260 configurations.

| $K_q$ | $m_\pi a$ | $m_\rho a$ | $m_N a$ |
|---|---|---|---|
| 0.1600 | 0.6873(24) | 0.8021(29) | 1.2900(60) |
| 0.1640 | 0.5080(29) | 0.6822(38) | 1.0738(80) |
| 0.1665 | 0.3674(39) | 0.6085(58) | 0.915(11) |

Table 2: Nucleon axial-vector matrix elements as a function of $K_q$.

| $K_q$ | $\Delta u_\text{conn.}$ | $\Delta d_\text{conn.}$ | $\Delta u_\text{disc.} = \Delta d_\text{disc.}$ | $\Delta s$ | | |
|---|---|---|---|---|---|---|
| | | | | $K_s = 0.160$ | 0.164 | 0.1665 |
| 0.1600 | 0.9071(92) | −0.2470(35) | −0.025(10) | −0.025(10) | −0.035(12) | −0.042(15) |
| 0.1640 | 0.839(19) | −0.2382(87) | −0.066(23) | −0.049(19) | −0.066(23) | −0.076(27) |
| 0.1665 | 0.818(39) | −0.231(23) | −0.093(54) | −0.068(35) | −0.084(46) | −0.093(54) |
| $K_c = 0.1694$ | 0.763(35) | −0.226(17) | −0.119(44) | −0.083(34) | −0.105(42) | −0.117(50) |

Table 3: Total contribution to $\Delta u$ and $\Delta d$, and $\Delta s$ at the physical strange quark mass corresponding to $K_s = 0.1648$. Flavor non-singlet coupling $F_A$ and $D_A$ are also listed.

| $K_q$ | $\Delta u_\text{total}$ | $\Delta d_\text{total}$ | $\Delta s$ | $F_A$ | $D_A$ |
|---|---|---|---|---|---|
| 0.1600 | 0.882(14) | −0.273(11) | −0.0374(89) | 0.4536(46) | 0.7011(37) |
| 0.1640 | 0.773(28) | −0.305(24) | −0.069(16) | 0.4196(93) | 0.6580(71) |
| 0.1665 | 0.715(64) | −0.326(56) | −0.087(33) | 0.409(20) | 0.641(16) |
| $K_c = 0.1694$ | 0.638(54) | −0.347(46) | −0.109(30) | 0.382(18) | 0.607(14) |

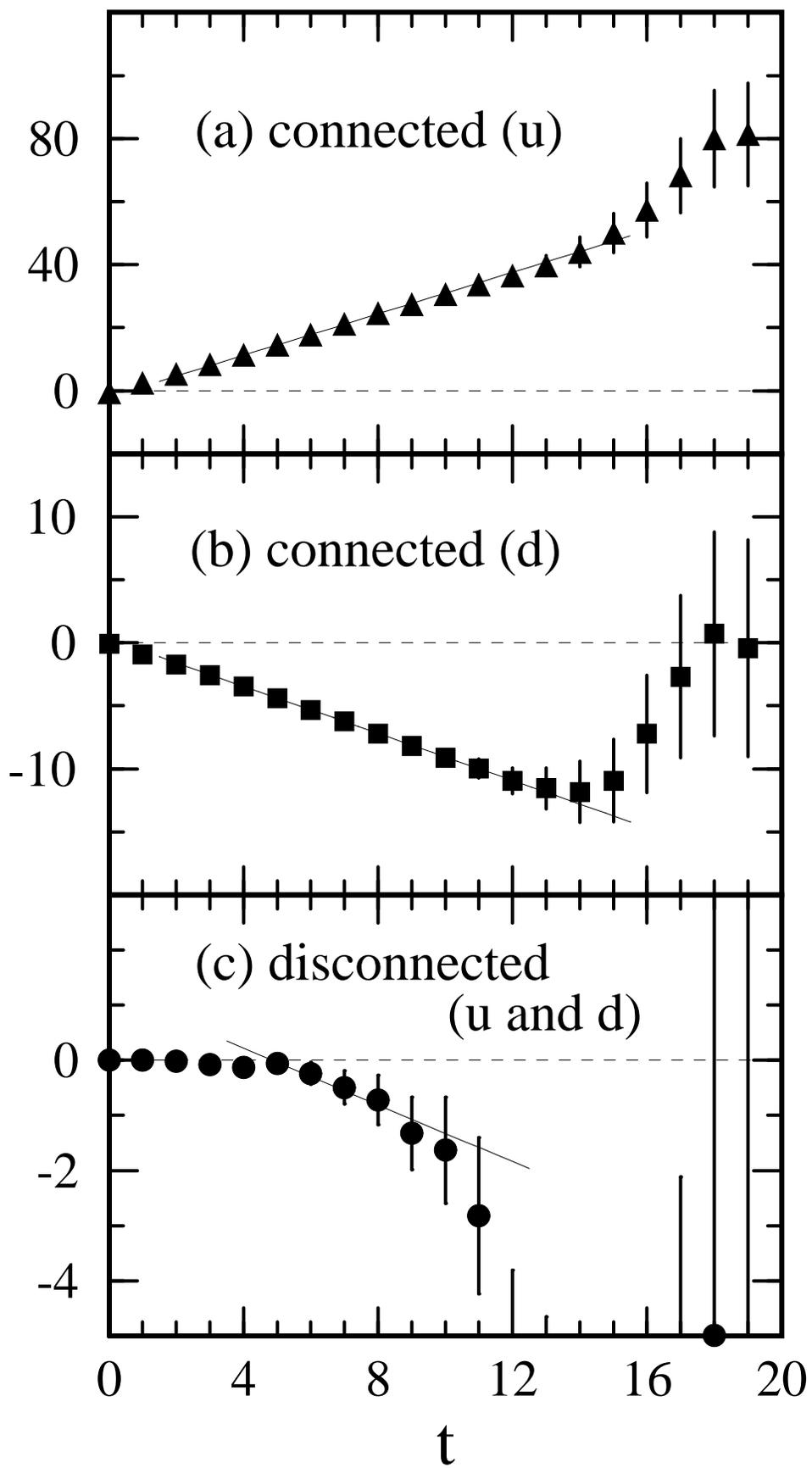

Fig.1

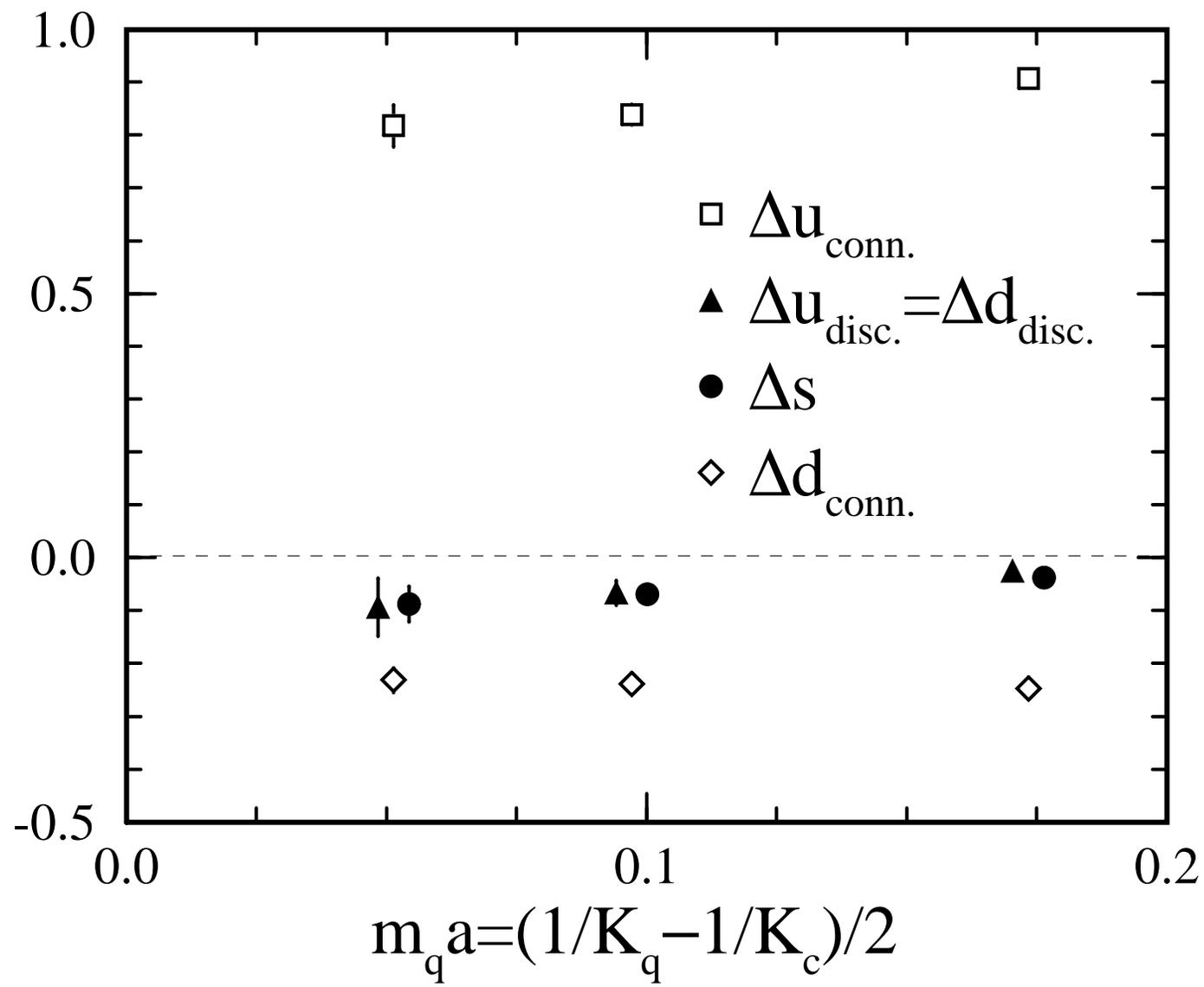

Fig.2